\documentclass[fleqn,usenatbib]{mnras}

\usepackage{newtxtext,newtxmath}
\usepackage{multirow}

\usepackage[T1]{fontenc}

\DeclareRobustCommand{\VAN}[3]{#2}
\let\VANthebibliography\thebibliography
\def\thebibliography{\DeclareRobustCommand{\VAN}[3]{##3}\VANthebibliography}

\usepackage{graphicx}	
\usepackage{amsmath}	
\usepackage{subcaption}
\usepackage{subcaption}

\title[TransientViT: CNN-ViT hybird model for real/bogus transient classification]{TransientViT: 
 A novel CNN - Vision Transformer hybrid real/bogus transient classifier for the Kilodegree Automatic Transient Survey}

\author[Chen et al.]{
Zhuoyang Chen,$^{1}$\thanks{E-mail: Placeholder}
Wenjie Zhou,$^{1}$
Guoyou Sun,$^{1}$
Mi Zhang,$^{1}$
Jiangao Ruan,$^{1}$
Jingyuan Zhao$^{1}$
\\
$^{1}$Xingming Observatory, Urumqi 830034, China\\
}

\date{Accepted XXX. Received YYY; in original form ZZZ}

\pubyear{2023}

\begin{document}
\label{firstpage}
\pagerange{\pageref{firstpage}--\pageref{lastpage}}
\maketitle

\begin{abstract}
The detection and analysis of transient astronomical sources is of great importance to understand their time evolution. Traditional pipelines identify transient sources from difference (\textit{D}) images derived by subtracting prior-observed reference images (\textit{R}) from new science images (\textit{N}), a process that involves extensive manual inspection. In this study, we present TransientViT, a hybrid convolutional neural network (CNN) - vision transformer (ViT) model to differentiate between transients and image artifacts for the Kilodegree Automatic Transient Survey (KATS). TransientViT utilizes CNNs to reduce the image resolution and 
 a hierarchical attention mechanism to model features globally. We propose a novel KATS-T 200K dataset that combines the difference images with both long- and short-term images, providing a temporally continuous, multidimensional dataset. Using this dataset as the input, TransientViT achieved a superior performance in comparison to other transformer- and CNN-based models, with an overall area under the curve (AUC) of 0.97 and an accuracy of 99.44\%. Ablation studies demonstrated the impact of different input channels, multi-input fusion methods, and cross-inference strategies on the model performance. As a final step, a voting-based ensemble to combine the inference results of three \textit{NRD} images further improved the model prediction reliability and robustness. This hybrid model will act as a crucial reference for future studies on real/bogus transient classification.
\end{abstract}




\section{Introduction}
Time-domain astronomy refers to the study of astronomical objects that change with time. In the quest for finding transient astronomical events, a multitude of large-scale optical sky surveys have been conducted, including the All-Sky Automated Survey for Supernovae (ASAS-SN; \citet{2014AAS...22323603S}), Dark Energy Survey (DES; \citet{dark2016dark}), Panoramic Survey Telescope and Rapid Response System (Pan-STARRS; \citet{kaiser2004pan}), Sloan Digital Sky Survey (SDSS; \citet{york2000sloan}), and Zwicky Transient Facility (ZTF; \citet{bellm2018zwicky}). These comprehensive all-sky surveys have enabled a thorough exploration of the optical sky by generating an overwhelming amount of image data, propelling astronomy into the currently emerging big data era (\citet{zhang2015astronomy}).

Conventional image processing pipelines locate and extract potential transient candidates, i.e. point sources, from the difference images constructed by subtracting a template image (taken at the time of observation) from the science image (\citet{cabrera2017deep}). Subsequently, candidates are classified as either true transients with real astrophysical significance or 'bogus' detections that are discarded as artifacts. This process typically requires extensive manual inspection. Given the ephemeral nature of transient events, their prompt and near real-time detection is crucial. A single night of observation can yield a plethora of bogus detections, rendering manual inspection of each candidate unfeasible. This has led to an increasing demand for rapid and accurate algorithms to distinguish true transients from artifacts.

Extensive efforts have been dedicated to integrating machine learning (ML) methods into image processing pipelines to facilitate the detection of transients. The pioneering work of \citet{romano2006supernova} allowed the application of support vector machines (SVMs; \citet{hearst1998support}) towards supernovae detection. \citet{bloom2012automating} and \citet{brink2013using} employed random forest (RF; \citet{breiman2001random}) classification algorithms to distinguish real and bogus transient candidates. \citet{goldstein2015automated}, \citet{du2015machine}, and \citet{wright2015machine} further compared several well-known traditional ML algorithms for real/bogus classification tasks and found RFs to exhibit superior performance compared to other mainstream approaches.

Deep learning (DL) methods involving convolutional neural networks (CNNs; \citet{fukushima1980neocognitron}), 
 frequently applied to diverse computer vision recognition tasks, are known to outperform conventional ML approaches (\citet{wang2019development}). \citet{wright2017transient} proposed an approach combining manual classification with CNN-based recognition for transient search. Furthermore, several flavors of CNN-based real/bogus classifiers have been put forward by \citet{andreoni2017mary}, \citet{cabrera2017deep}, \citet{duev2019real}, \citet{hosenie2021meercrab}, \citet{takahashi2022deep}, and \citet{acero2022s}. \citet{yin2021supernovae} extended the framework by developing a fully convolutional one-stage (FCOS; \citet{tian2019fcos}) algorithm for supernova detection. Despite their effectiveness, CNNs face limitations in describing low-level features beyond the effective receptive fields. Therefore, it is not conducive to making full use of the context information to capture the features of images. Stacking deeper convolutional layers aids in extracting higher levels of image features, but substantially increases the computational costs (\citet{chen2022fast}).

To address the aforementioned issue, we propose a hybrid CNN - vision transformer (ViT; \citet{dosovitskiy2020image}) model, named TransientViT, for real/bogus transient classification. ViTs can facilitate classification as they utilize a self-attention mechanism that enables global information integration, rather than being limited to local information specific to individual transients. CNN-ViT hybrid models are equipped with the locality of CNNs as well as the global connectivity of ViTs (\citet{MANZARI2023106791}). Additionally, we emphasized the reduction of computational and inference time costs for the proposed real/bogus classifier, considering the high computational demand of the original ViT model.

The remainder of this paper is structured as follows: In Section 2, we introduce our dataset obtained from the Kilodegree Automatic Transient Survey (KATS) telescope array. Section 3 describes the overall architecture of the proposed TransientViT model. In Section 4, we present our experimental results and compare them to other mainstream ViT- and CNN-based models. Finally, we summarize and present our conclusions in Section 5.

\section{Dataset}
\subsection{KATS-T 200K}
We propose a novel KATS-T 200K dataset, consisting of nine images per set of observation data, encompassing both long- and short-term images. The images in our dataset were acquired from KATS conducted at the Xingming Observatory, Xinjiang, China. KATS comprises an array of six 0.28 m Rowe-Ackermann Schmidt Astrograph (RASA) telescopes, with a field of view of 6.7 $\times$ 6.6 square degrees. For the transient survey, 30 s exposure images are taken without the use of filters, yielding a typical limiting magnitude of 19 mag. The dataset consists of transient candidate detections captured by KATS from April 1, 2023 to July 29, 2023. From a total of 201,358 samples, 561 confirmed transients were reported in the Transient Name Server (TNS)\footnote{https://www.wis-tns.org}, along with 200,797 bogus detections. The dataset was split into training, validation, and test sets as shown in Table~\ref{tab:dataset}. 

\begin{figure}
	\includegraphics[width=\columnwidth]{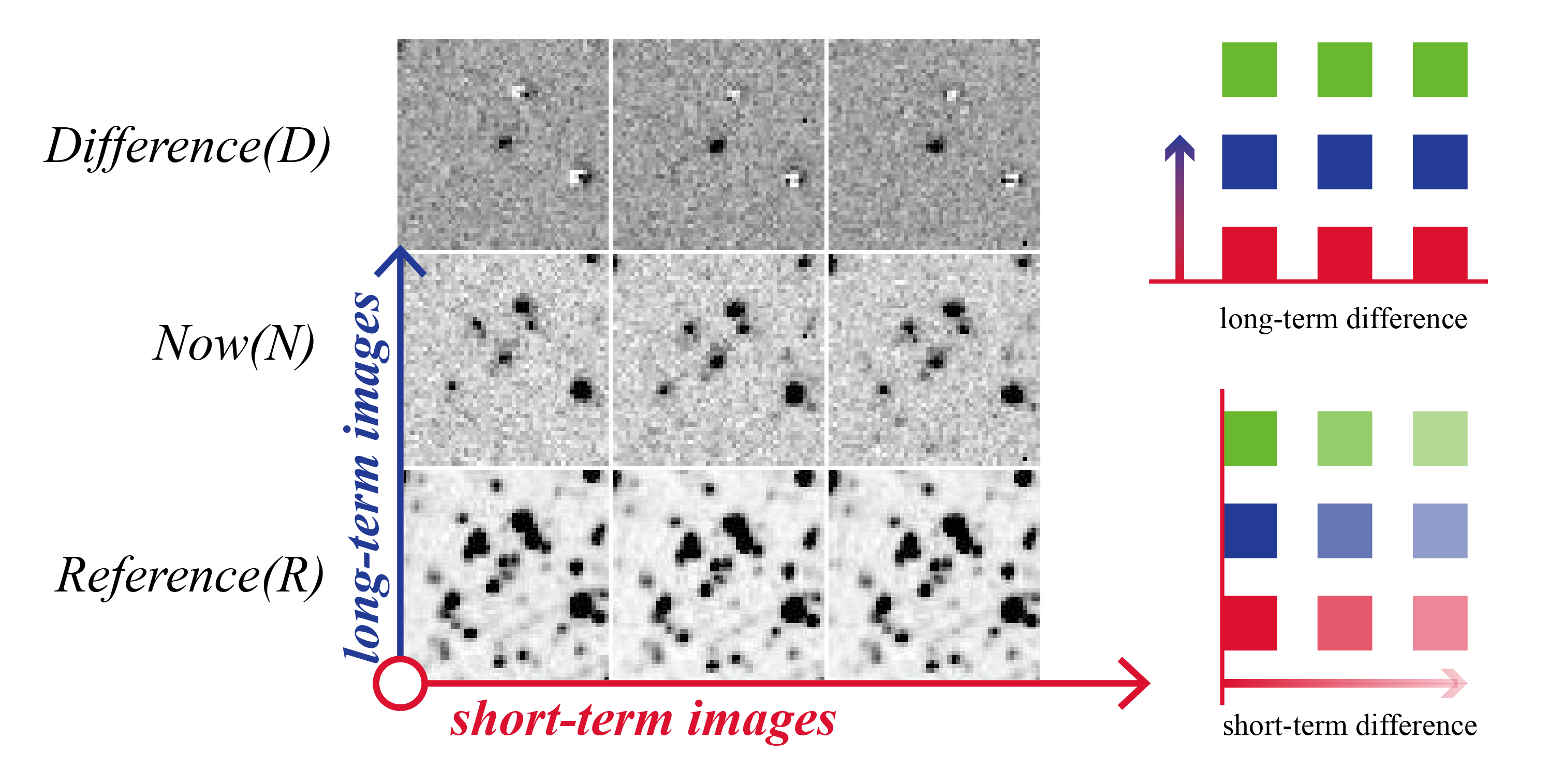}
	\caption{KATS-T 200K sample images.}
	\label{fig:sample}
\end{figure}

Transients are identified within difference images derived by subtracting the new image, taken at the time of observation, from the reference image, captured on a prior date. The proposed KATS-T 200K dataset incorporates these commonly used difference images as well as long-term images (observed over larger intervals) and short-term images (observed over shorter intervals). In Fig.~\ref{fig:sample}, the long-term images are presented in a vertical sequence (top-to-bottom: difference, new, and reference image), and the short-term images, captured on the same day at varying intervals, are presented in a horizontal sequence. The combination of long- and short-term images allows for a more continuous temporal sequence, providing multidimensional data for real/bogus classification.

\begin{table}
	\centering
	\caption{Number of images in the training, validation, and test datasets.}
	\label{tab:dataset}
	
	\begin{tabular}{lccr} 
		\hline
		Split & Real & Bogus\\
		\hline
		Training & 401 & 197,430\\
		Validation & 80 & 560\\
		Testing & 80 & 2,807\\
		\hline
	\end{tabular}
	
\end{table}

\subsection{Data preprocessing}

\begin{figure}
	\includegraphics[width=\columnwidth]{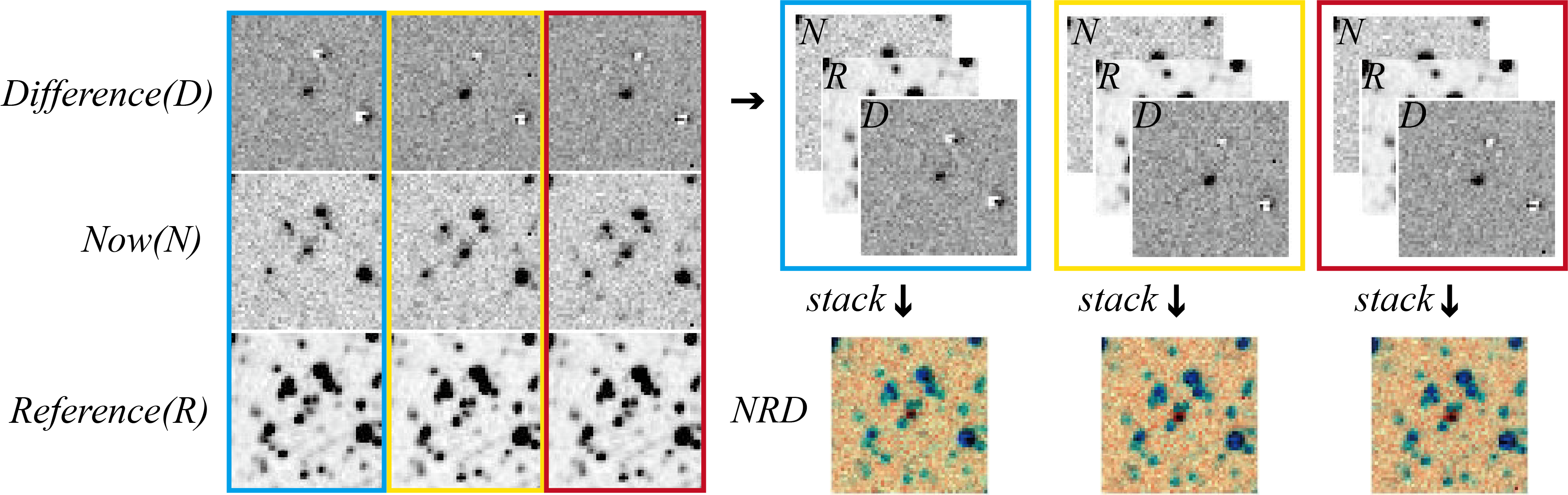}
	\caption{Data preprocessing for the KATS-T 200K dataset.}
	\label{fig:preprocess}
\end{figure}

The preprocessing procedure for the KATS-T 200K dataset is shown in Fig.~\ref{fig:preprocess}. The vertical axis represents the long-term difference (\textit{D}), new (\textit{N}), and reference (\textit{R}) images, while the horizontal axis represents the images at varying times within the same day with shorter intervals. The image is first divided into three short-term data segments horizontally. For each segment, the \textit{N}, \textit{R}, and \textit{D} images are stacked (in that order) to generate an \textit{NRD} three-channel image (\citet{meercrab}). In other words, each sample is processed into three \textit{NRD} images with temporal information.

\begin{figure*}
	\includegraphics[width=\textwidth]{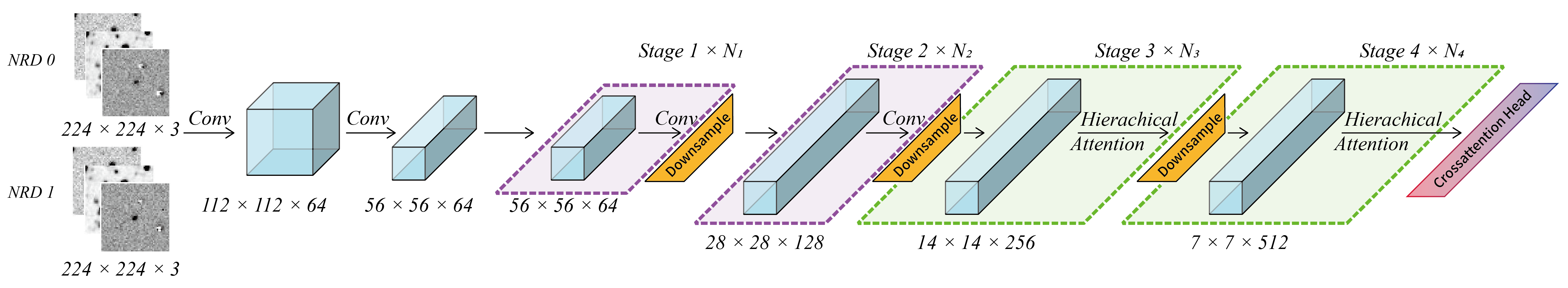}
	\caption{Architecture of the TransientViT model.}
	\label{fig:network}
\end{figure*}

\section{Methods}

\subsection{Model architecture}

The structure of the TransientViT model (Fig.~\ref{fig:network}) combines the benefits of the fast local representation learning of the CNNs and the global modeling properties of the ViTs (\citet{dosovitskiy2021image}). The architecture of TransientViT is organized into three components. The first segment comprises the initial convolutional layers, which extract features from high-resolution feature maps. The second segment utilizes hierarchical attention layers to perform spatial reasoning on the entire feature map and models the global information. The third segment employs adaptive cross-attention to fuse the feature representations of the different short-term segments across the image. TransientViT enhances the efficiency of the model by employing convolutional layers to decrease the image resolution and utilizes a hierarchical attention mechanism to model the global feature map. This approach effectively reduces the computational burden of the attention mechanism, thereby reducing the inference time.
\begin{figure*}
	\includegraphics[width=\textwidth]{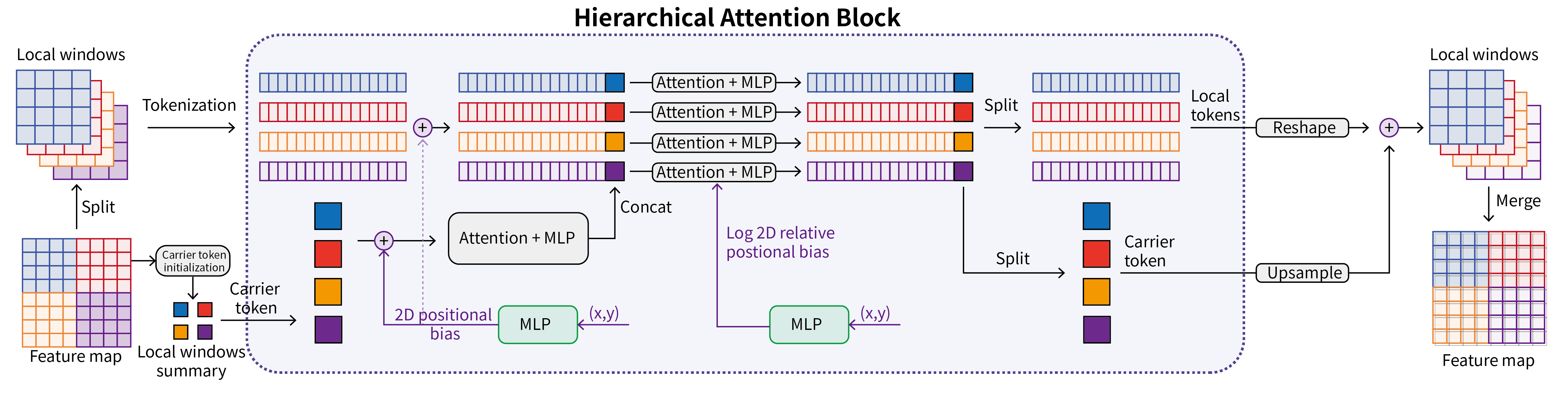}
	\caption{Structure of the hierarchical attention block.}
	\label{fig:hat}
\end{figure*}
\subsection{TransientViT Components}
\subsubsection{Stem}

An input image is divided into non-overlapping patches using two consecutive 3×3 convolutions with a stride of 2. These patches are then transformed into \textit{D}-dimensional embeddings. After each convolution, batch normalization (BN) and rectified linear unit (ReLU) activation (\citet{relu}) are applied to the embeddings.

\subsubsection{Downsample Blocks}

TransientViT utilizes a hierarchical structure, which implements downsampling before the stage layer to decrease the feature map resolution by a factor of 2.

\subsubsection{Convolutional Blocks}

The \textit{Conv} blocks (Fig.~\ref{fig:network}) constitute stages 1 and 2, both comprised of residual convolutional blocks. The output of the \textit{Conv} block can be expressed as (\citet{gelu})

\begin{equation}
    \hat{x}={\rm GELU}[{\rm BN}{(Conv_{3\times3}(x))}],
    \label{eq:GELU}
\end{equation}

\begin{equation}
    x={\rm BN}(Conv_{3\times3}(\hat{x})) + x
    \label{eq:GELU2}.
\end{equation}

\subsubsection{Hierarchical Attention}

The hierarchical attention structure is incorporated in stages 3 and 4, which was first proposed for the FasterViT model (\citet{fastervit}). It decomposes the quadratic time complexity of global self-attention into multiple simpler attention mechanisms, effectively mitigating computational overhead. The approach initiates by adopting the use of local windows, as employed for the Swin Transformer (\citet{swin}). Subsequently, carrier tokens (CTs) are introduced to summarize elements for the entire local window. Global information is summarized and propagated by applying the first attention block to the CTs. To ensure localized access, the local window tokens and CTs are concatenated, allowing each local window to exclusively access its corresponding set of CTs. By applying self-attention to the concatenated tokens, efficient exchange of both local and global information is enabled while minimizing computational costs. The concept of hierarchical attention was formulated by alternating between sub-global (CTs) and local (windowed) self-attention. Conceptually, CTs can be further grouped into windows, featuring a higher order of CTs.
\begin{figure*}
	\includegraphics[width=\textwidth]{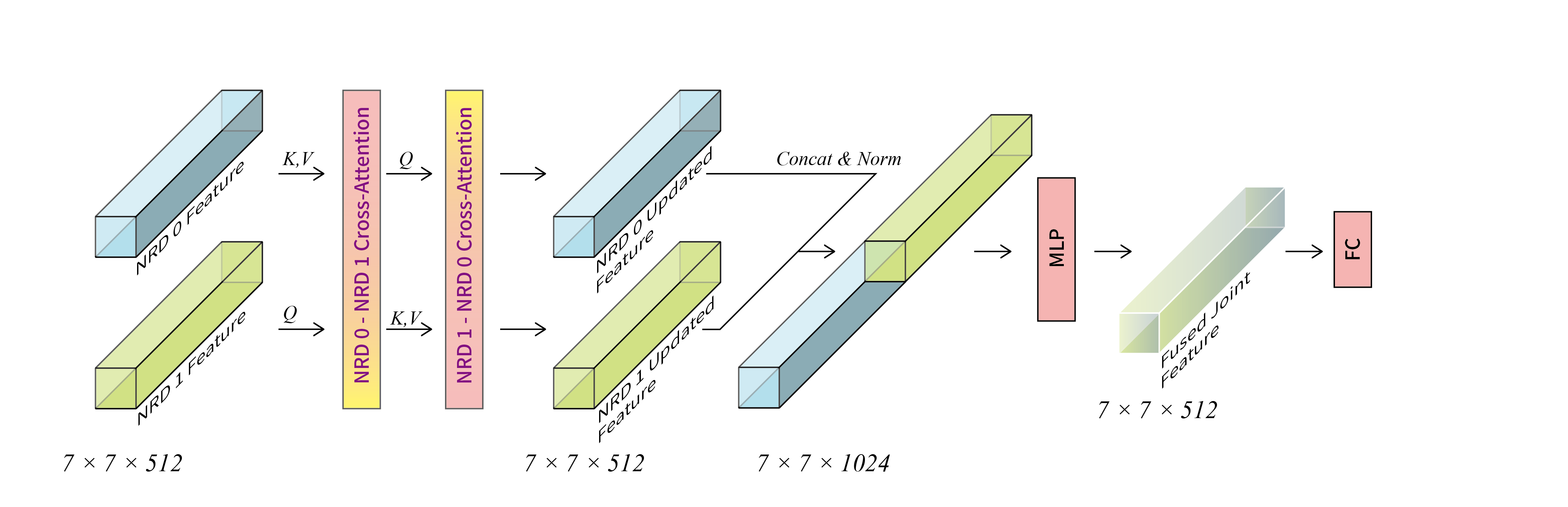}
	\caption{Adaptive cross-attention head architecture.}
	\label{fig:crossattention}
\end{figure*}

\begin{figure*}
	\includegraphics[width=\textwidth]{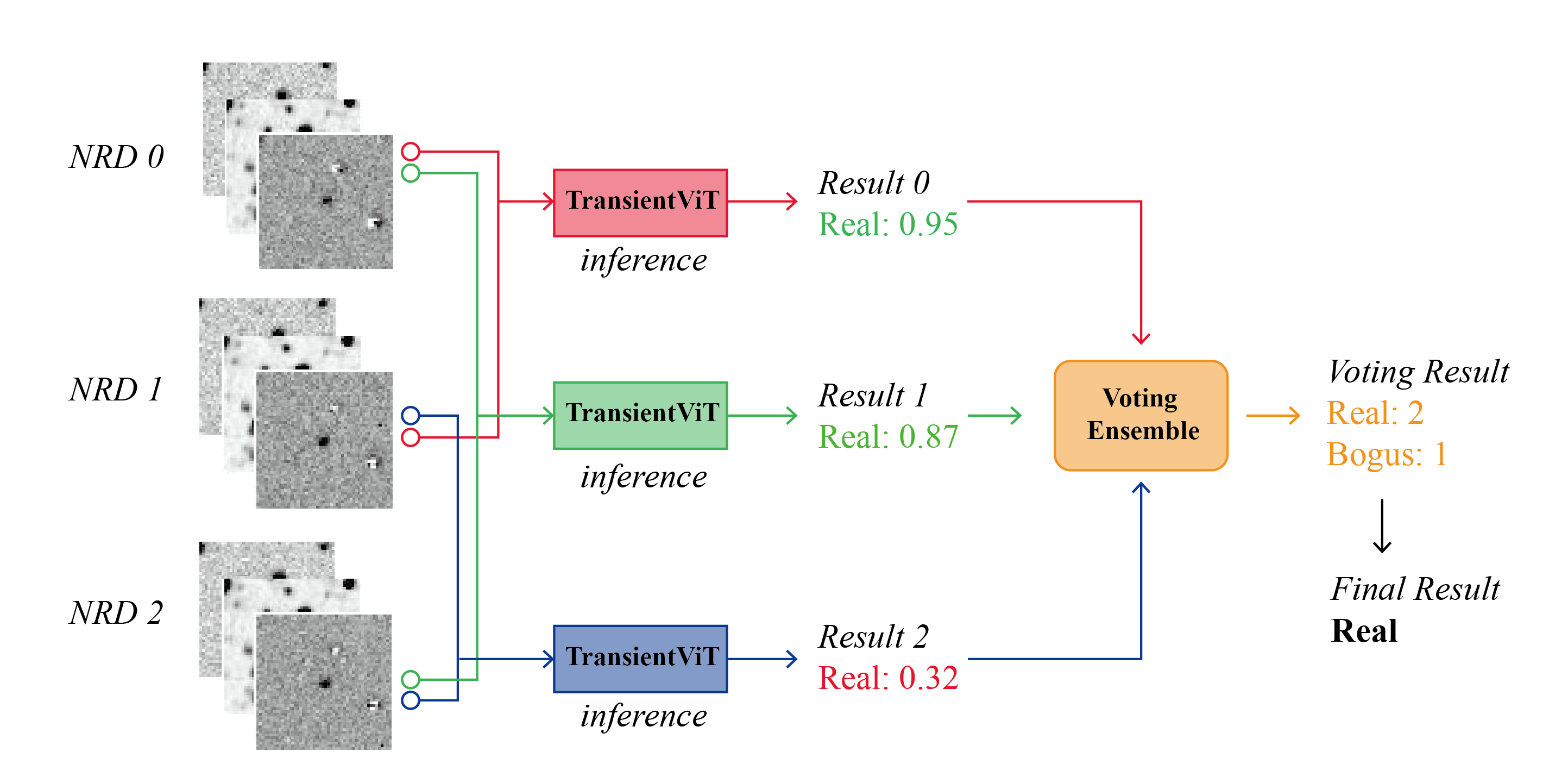}
	\caption{Cross-inference process.}
	\label{fig:crossinference}
\end{figure*}
\subsubsection{Adaptive Cross-Attention Head}
\label{sec:crossattn}
The feature information extracted from stage 4 is directed into the adaptive cross-attention head to facilitate feature fusion and classification (Fig.~\ref{fig:crossattention}). By processing three \textit{NRD} images, three corresponding feature maps are generated. The adaptive cross-attention head dynamically selects two feature maps to perform cross-attention computation. Subsequently, the selected feature maps are concatenated and subjected to layer normalization (LN) expressed as (\citet{ln})

\begin{equation}
	\hat{x}_{tc} = \gamma_c\frac{x_{tc} - \mu_t^{ln}}{\sqrt{(\sigma_{t}^{ln})^2+\epsilon}} + \beta_c,
	\label{eq:ln}
\end{equation}

\noindent where \textit{t} and \textit{c} are 
 the indices and embeddings of a token, respectively,  $\epsilon$ is a small positive constant to avoid a zero denominator, $\gamma_c$ and $\beta_c$ are two learnable parameters in the affine transformation, and the LN normalization constants, $\mu_t^{ln}$ and ${(\sigma^2)_t^{ln}}$, can be expressed as

\begin{equation}
    \mu_{t}^{ln} = \frac{1}{C}\sum_{c=1}^{C}x_{tc}, 
\end{equation}
\begin{equation}
    \sigma_{t}^{ln} = \sqrt{\frac{1}{C}\sum_{c=1}^{C}(x_{tc} - \mu_{t}^{ln})^2}.
    \label{eq:mu}
\end{equation}

Finally, we apply the multi-layer perceptron (MLP; \citet{mlp}) and fully connected (FC) layers. TransientViT effectively models the spatial information in the \textit{NRD} images and captures temporal information from short-term data segments, amplifying the capability of the model to distinguish between real and bogus transient candidate detections.

\subsubsection{Cross Inference}

After preprocessing, the three \textit{NRD} segments from the KATS-T 200K dataset were directed into the TransientViT model for inference, generating individual prediction results. Ultimately, a voting-based ensemble was applied to the individual inferences to obtain the final classification result (Fig.~\ref{fig:crossattention}).

\section{Results}

\subsection{Implementation details}

The TransientViT was implemented using PyTorch 1.12.0 (\citet{NEURIPS2019_9015}) on Python 3.7. To train the TransientViT model, we leveraged 8 Nvidia GeForce RTX 3090 GPUs, each equipped with a VRAM of 24GB. During training, we employ the AdamW (\citet{adamw}) optimizer with a low learning rate (lr = 0.0001) and a batch size of 32. The learning rate scheduler followed a cosine decay strategy (\citet{cosine}). Our model utilized the cross-entropy loss function expressed as

\begin{equation}
    {\rm CE}=-{(y\log(p) + (1 - y)\log(1 - p))},
    \label{eq:ce}
\end{equation} 

\noindent where \textit{y} is the binary indicator for a class and \textit{p} denotes the probability assigned to that class.


\begin{table}
	\centering
        \caption{TransientViT training settings.}
	\label{tab:train detail}
	\begin{tabular}{lc} 
		\hline
		Configuration & Parameters  \\
		\hline 
		pretrain & ImageNet-1k (224×224) \\
		optimizer & AdamW  \\
		base learning rate & 1e-4  \\
		warmup learning rate & 1e-5 \\
		weight decay & 0.1  \\
		batch size & 256 \\
		training epochs & 200 \\
		learning rate schedule & cosine decay \\
		warmup epochs & 20 \\
		randaugment & (9, 0.5) \\
		mixup & None \\
		cutmix & None \\
		random erasing & 0.2 \\
		label smoothing & 0.1 \\
		
		\hline
	\end{tabular}
	
\end{table}

We employ offline data augmentation to mitigate the risk of overfitting, thus avoiding poor generalization of the model for the test set. Our primary data augmentation techniques encompassed color jitter, RandAugment (magnitude 9; \citet{randaug}), random horizontal flip, and random vertical flip. These offline data augmentation strategies noticeably amplify the variations in the training dataset.

\subsection{Metrics}

We considered real transients as positives (\textit{p}) and bogus detections as negatives (\textit{n}). The probability generated by TransientViT to indicate whether the source at the center is to be classified as a bogus or real transient is denoted as \textit{P}. In order to arrive at a definitive decision, it is imperative to establish a probability threshold. Our primary objective was to minimize false positives (FP), while maintaining the lowest feasible level for false negatives (FN; ideally below 5\% of the total number of candidates). Beyond assessing the classification performance, we also incorporated additional metrics as explained below:

\begin{itemize}
    \item Precision (Prec): This metric calculates the number of real transients among all the objects classified as transients by TransientViT. A high precision score indicates that the model is consistently accurate in predicting the positive class representing real transients. However, the dataset used in this study was highly imbalanced. Thus, precision might not serve as a reliable indicator of model performance within this context. The metric is defined as
    \begin{equation}
        {\rm Prec}=\frac{{\rm TP}}{\rm (TP+FP)}.
	\label{eq:prec}
    \end{equation}

    \item Recall: This metric gauges the quantity of the accurately classified transients within the dataset. A high recall score signifies the adeptness of the model in detecting a substantial proportion of transients. This metric is defined as

    \begin{equation}
	{\rm Recall}=\frac{{\rm TP}}{\rm (TP+FN)}.
	\label{eq:recall}
    \end{equation}

\item Receiver Operating Characteristic (ROC) Curve: This metric measures the relationship of the true positive rate (TPR) with the false positive rate (FPR). An ideal model would exhibit a vertical line at $x ={\rm FPR}=0$ and a horizontal line at $y={\rm TPR}=1$. This metric provides a visual representation of the overall performance of the model, allowing a comprehensive evaluation of its efficacy. Here, FPR and TPR are defined as
\begin{equation}
	{\rm FPR}=\frac{{\rm FP}}{{\rm TN+FP}}
	\label{eq:fpr}
\end{equation}

\begin{equation}
	{\rm TPR}=\frac{{\rm TP}}{{\rm TP+FN}}
	\label{eq:tpr}
\end{equation}

\item Precision-Recall (P-R) curve: This metric demonstrates the model performance for classifying the $p$ class (real transients).

\item Area Under the Curve (AUC): This metric measures the overall performance of a binary classifier. The AUC value is within the range [0.5–1.0], where the minimum value represents the performance of a random classifier and the maximum value corresponds to a perfect classifier (i.e., with a classification error rate of zero). (\citet{Melo2013})
\begin{equation}
	{\rm AUC}=\int{\rm TPR}  d({\rm FPR})
	\label{eq:auc}
\end{equation}

\end{itemize}

\begin{table*}
	\centering
	\caption{Detailed comparison of different classification models.}
	\label{tab:backbone}
	\begin{tabular*}{\linewidth}{@{\extracolsep{\fill}} l|cccccc } 
		\hline
		Model & Image Size & Parameters (M) & Validation Accuracy & Validation AUC & Test Accuracy & Test AUC \\
		\hline
		ResNet-50 & 224 & 23.51 & 97.93 & 96.05 & 98.81 & 97.84\\
		ResNet-101 & 224 & 42.5 & 95.20 & 95.08 & 97.95 & 94.07\\
		ConvNeXt v2 nano & 224 & 14.98 & 97.87 & 95.99 & 99.19 & 96.77\\
		ConvNeXt v2 tiny & 224 & 27.87 & 98.40 & 96.39 & 99.35 & 90.68\\
		Swin Transformer v2 tiny & 256 & 28.35 & 98.52 & 98.10 & 99.34 & 96.46\\
		Swin Transformer v2 small & 224 & 49.73 & 98.28 & 97.92 & 99.14 & 95.89\\
		EfficientViT b1 & 224 & 7.50 & 95.50 & 96.28 & 98.56 & 93.86\\
		EfficientViT b2 & 224 & 21.77 & 95.56 & 95.51 & 98.60 & 94.27\\
		EfficientViT b3 & 224 & 46.09 & 96.21 & 96.14 & 98.82 & 95.76\\
		\textbf{TransientViT (ours)} & 224 & 30.89 & \textbf{98.58} & \textbf{98.72} & 99.44 & \textbf{97.88}\\
		\hline
	\end{tabular*}
	
\end{table*}
\subsection{Experiment results}

\begin{figure}
	\includegraphics[width=\columnwidth]{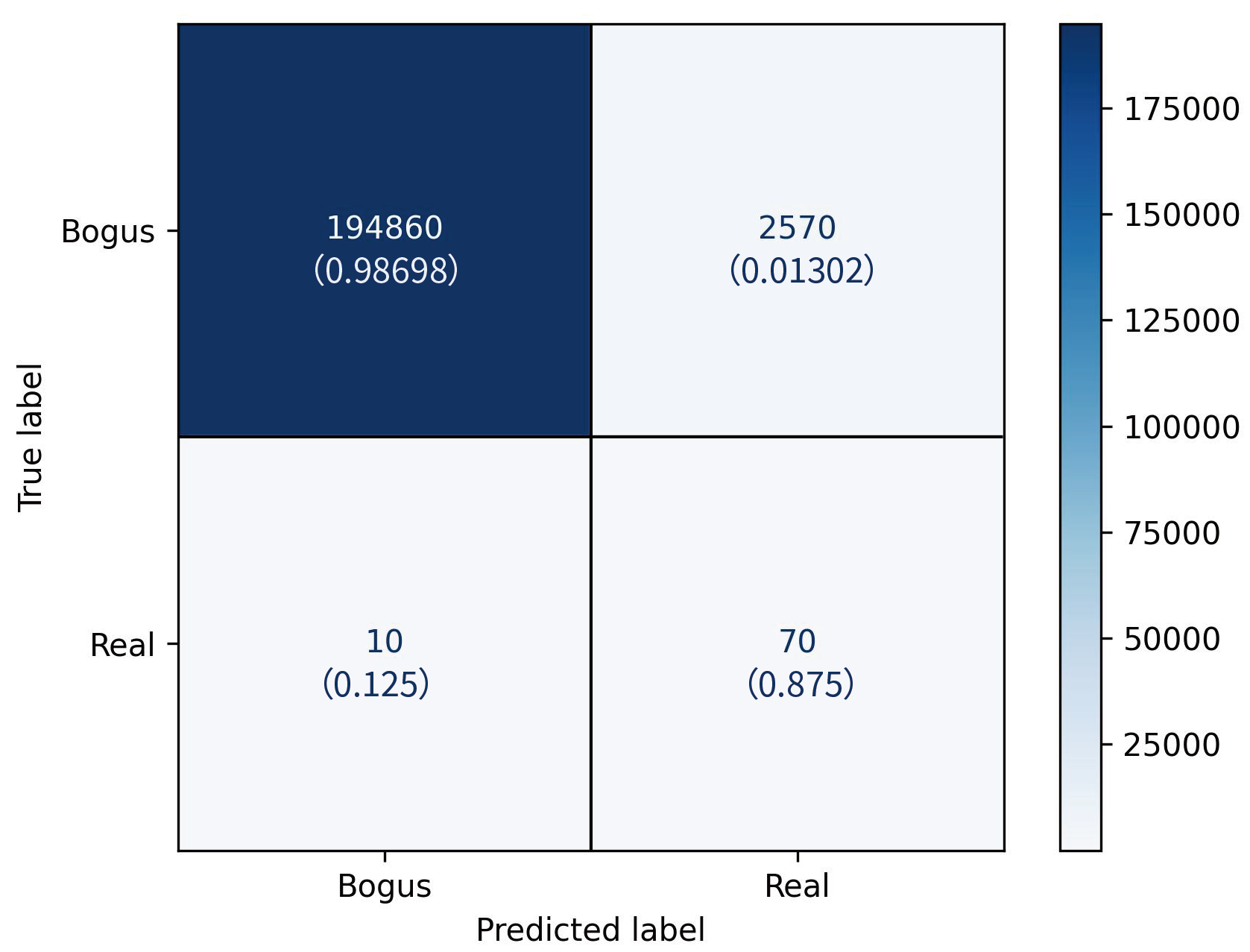}
	\caption{Confusion matrix for TransientViT. The predicted labels for all the classifiers are obtained using a threshold of 0.5. The values within each row are normalized. The numbers presented outside the parentheses represent the raw counts. }
	\label{fig:confusion matrix}
\end{figure}

Based on the predicted and actual classes, we constructed a confusion matrix to have an overview of the classification results (Fig.~\ref{fig:confusion matrix}). The confusion matrix displays the fraction of correctly classified candidates as the diagonal elements (TP and TN). The off-diagonal values show the misclassified examples (FP and FN). The ROC curve is shown in Fig. 9b, with an area under the curve (AUC) of 0.97. The loss and accuracy curves during the training and validation processes are shown in Fig. 10. The model converged at the 50th epoch during training. The validation loss demonstrated a tendency to stabilize in the later phases of training, implying that the model did not significantly overfit.

\begin{figure}
	\includegraphics[width=\columnwidth]{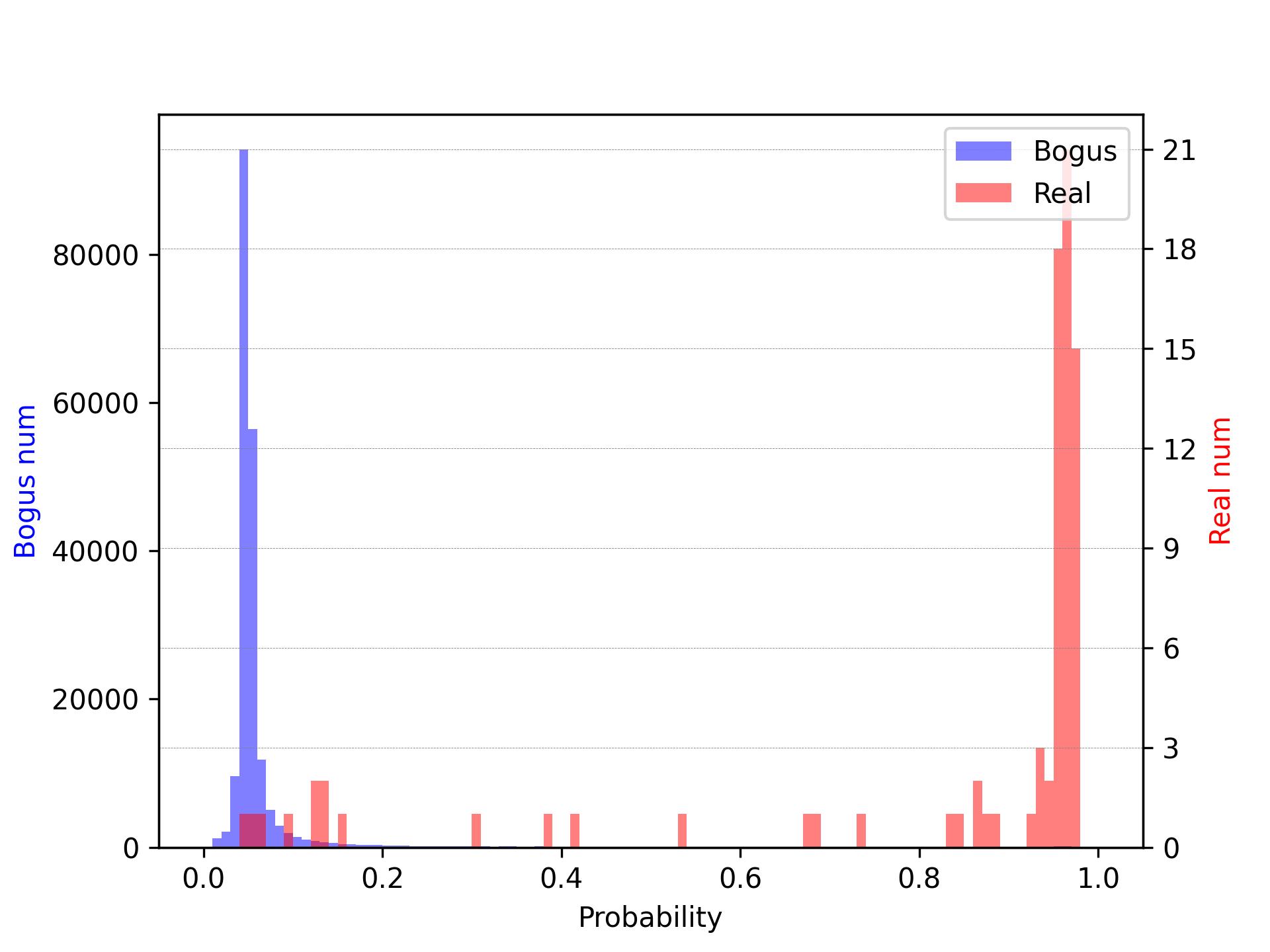}
	\caption{Real (red) / bogus (blue) probability distribution.}
	\label{fig:hist}
\end{figure}

\begin{figure}
	
\begin{subfigure}{0.5\columnwidth}
	\includegraphics[width=\textwidth]{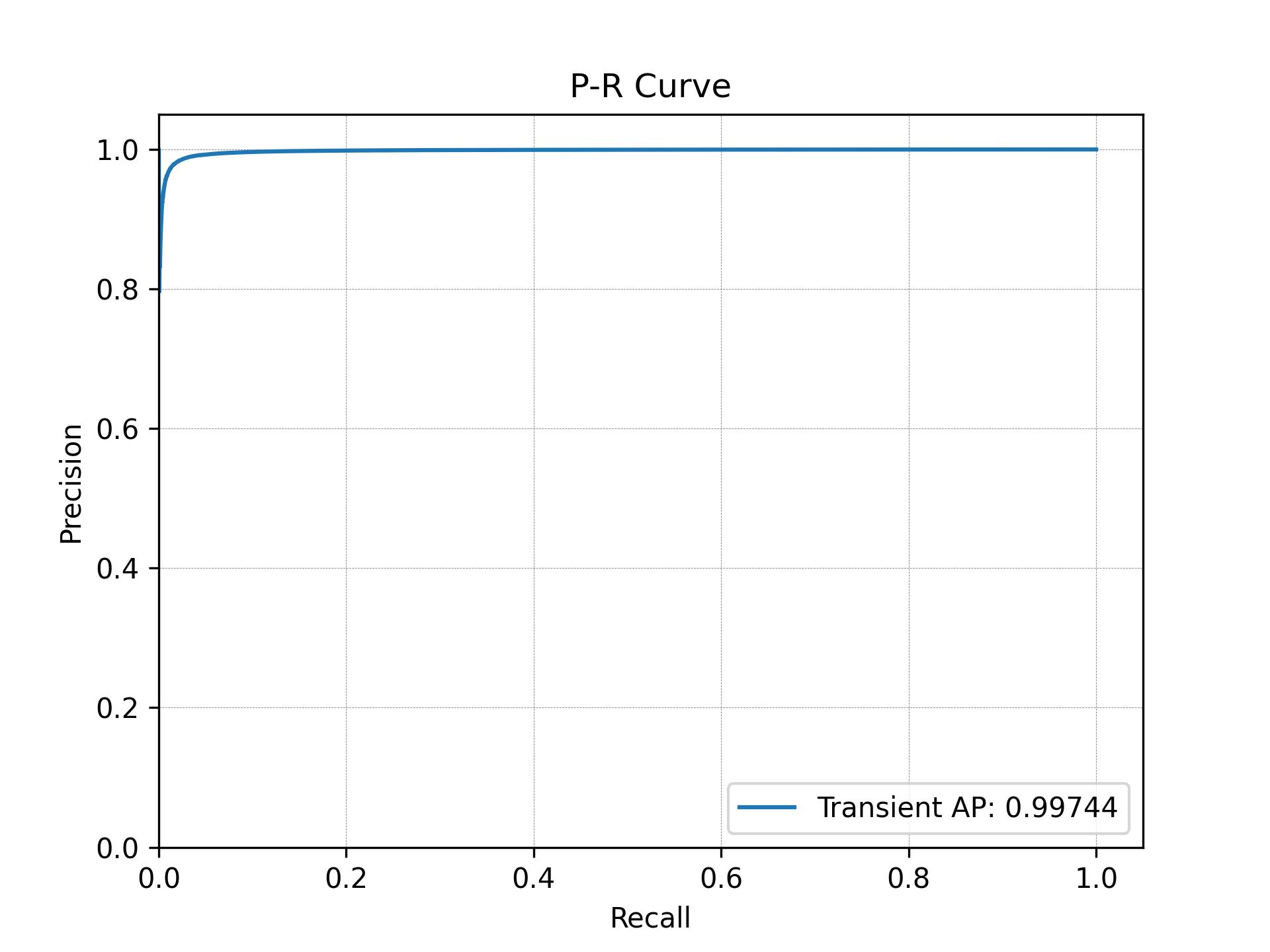}
	
	\label{fig:pr}
\end{subfigure}
\hfill
\begin{subfigure}{0.5\columnwidth}
	\includegraphics[width=\textwidth]{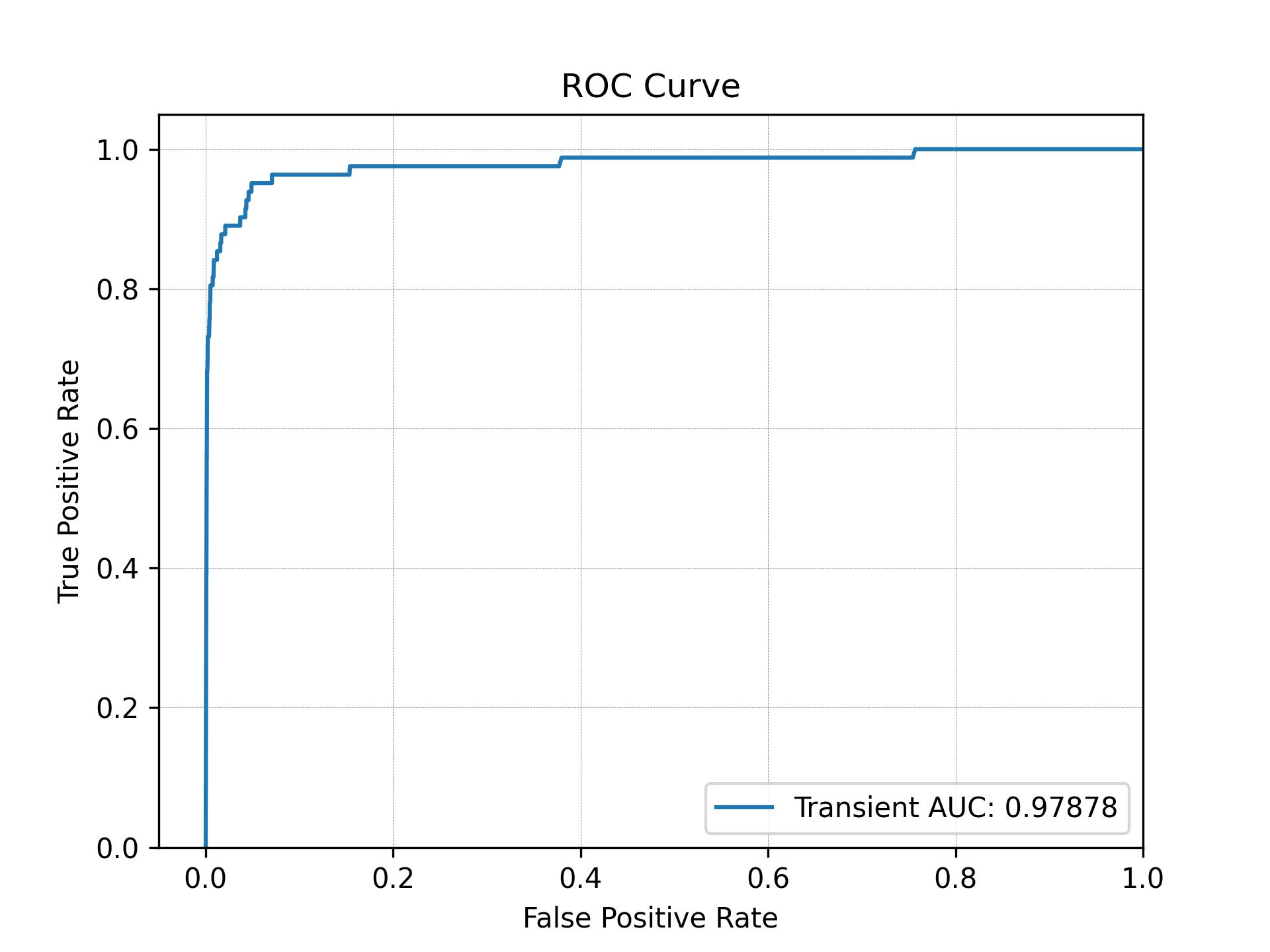}

\end{subfigure}
\label{fig:roc}
\caption{ (a) P-R and (b) ROC curves for TransientViT.}

\end{figure}

\begin{figure}
	
	\begin{subfigure}{0.5\columnwidth}
		\includegraphics[width=\textwidth]{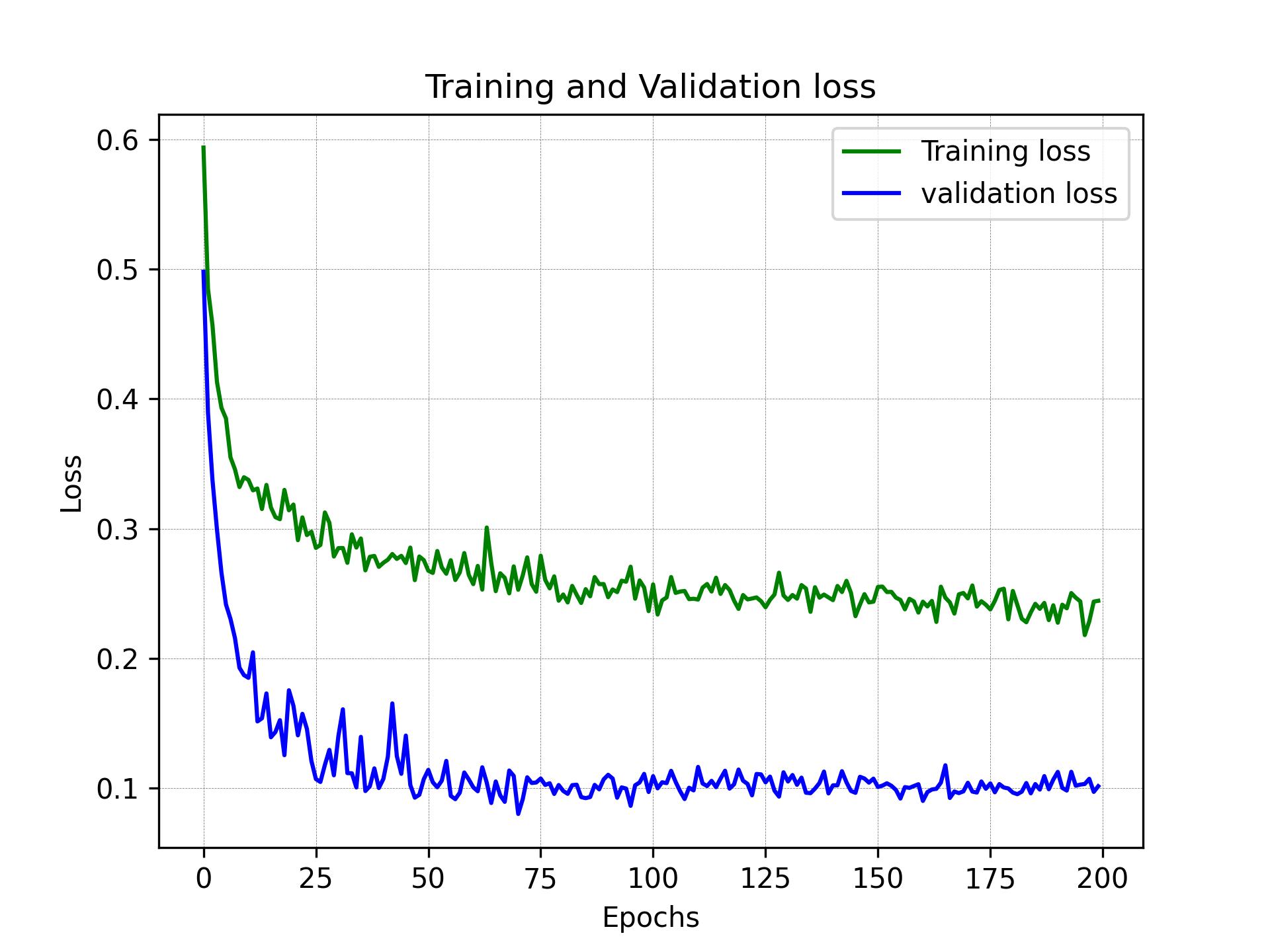}
		
		\label{fig:loss}
	\end{subfigure}
	\hfill
	\begin{subfigure}{0.5\columnwidth}
		\includegraphics[width=\textwidth]{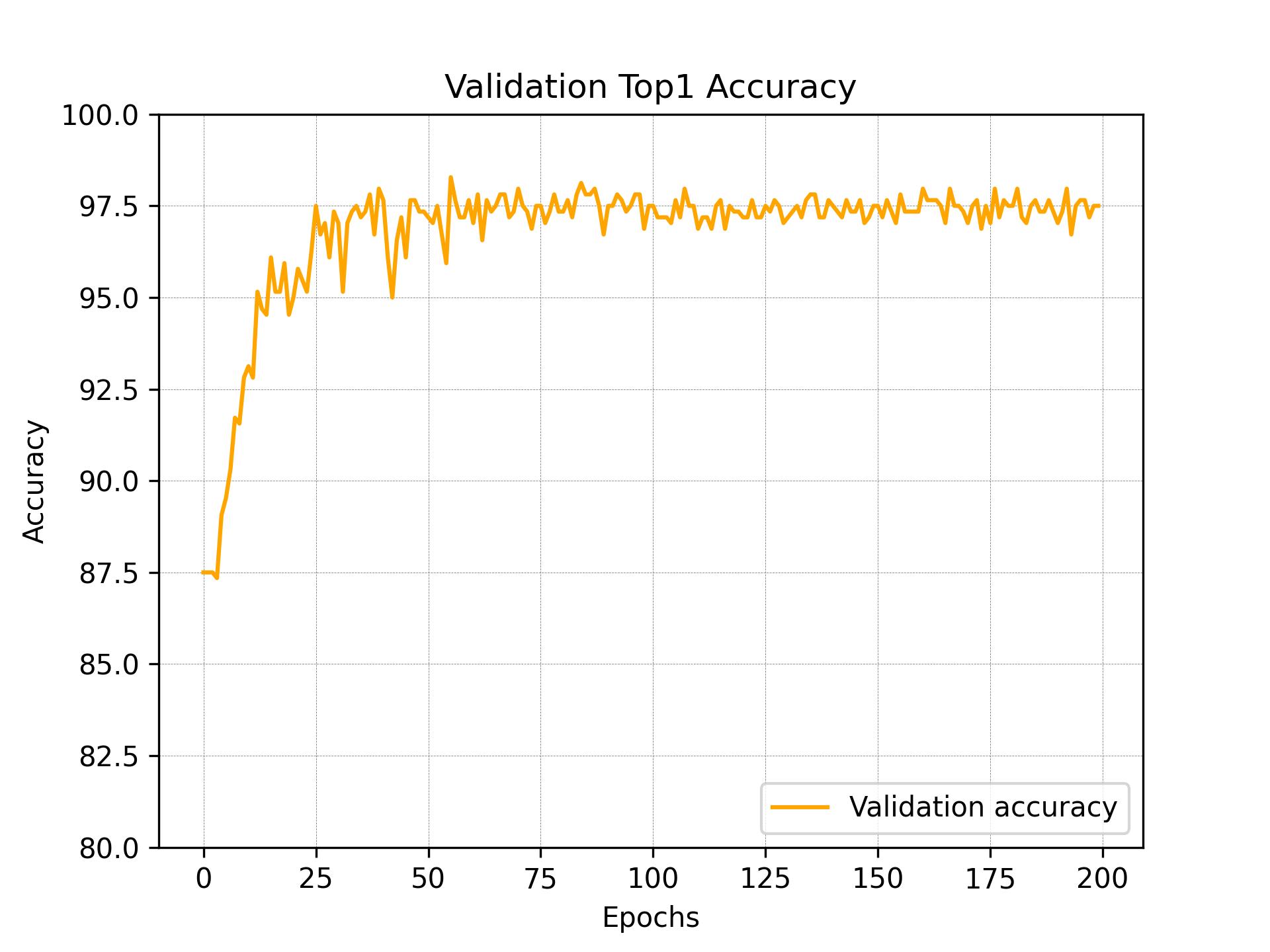}
		
		\label{fig:acc}
	\end{subfigure}
	
	\caption{(a) Loss for the training and validation sets and (b) accuracy for the validation for TransientViT.}
	
\end{figure}

\subsection{Ablation study}

\subsubsection{Backbone}
We conducted a performance evaluation of multiple classification models trained on the KATS-T 200K dataset, and compared them with the proposed TransientViT model (Table.~\ref{tab:backbone}). For similar parameter count, TransientViT exhibited superior performance compared to other transformer-based models, such as EfficientViT and Swin Transformer, as well as CNN-based models, such as ConvNeXt and ResNet, for transient classification. It is important to note that,\ due to the distinct characteristics of TransientViT, we adjusted the adaptive cross-attention head to a conventional MLP head during the backbone comparison, facilitating a standard single-image input.

\subsubsection{Image channel}

We conducted experiments to assess the training performance of different image channels using TransientViT (Table~\ref{tab:channel}). Training with the \textit{NRD} images yielded superior metrics in comparison to using a single Diff channel.

\begin{table}
	\centering
	\caption{Accuracy and AUC for different channel images employed in TransientViT.}
	\label{tab:channel}
	\begin{tabular}{lcccc} 
		\hline
		Model & Channel  & Test Accuracy & Test AUC \\
		\hline
		\multirow{2}{*}{TransientViT} & Diff & 99.23 & 97.36 \\
		& NRD & \textbf{99.24} & \textbf{97.90} \\
		\hline
	\end{tabular}
	
\end{table}

\subsubsection{Multi-input fusion}

We conducted several sets of experiments for multiple input images and explored various fusion methods for TransientViT (Table.~\ref{tab:feature}), namely:

\begin{enumerate}
    \item \textbf{SuperImage:} We directly utilized grayscale images with a patch size of 3$\times$3 as input for the model.
    
    \item \textbf{Feature Concatenation:} We concatenated the features extracted from two image segments using TransientViT, along the channel dimension, resulting in an increase in feature depth from 1\textit{C} to 2\textit{C}. Subsequently, the concatenated feature representation was utilized as the input for the conventional MLP head.

    \item \textbf{Feature Addition:} We performed element-wise addition of the features extracted from two image segments, while preserving the original feature depth.

    \item  \textbf{Adaptive cross-attention:} See Section ~\ref{sec:crossattn} for details.
\end{enumerate}

\begin{table}
	\centering
	\caption{Accuracy and AUC for different multi-input processing methods.}
	\label{tab:feature}
	\begin{tabular}{lccc} 
		\hline
		Model &  Multi Input Processing & Test Accuracy & Test AUC\\
		\hline
		\multirow{5}{*}{TransientViT} & SuperImage & 97.18 & 96.55 \\
		& Feature Concatenation & 98.43 & 97.61 \\
		& Feature Addition & 98.12 & 97.11 \\
		& Adaptive Cross-Attention & \textbf{98.66} & \textbf{97.87} \\
		\hline
	\end{tabular}
	
\end{table}

\subsubsection{Cross-inference}

We evaluated the performance of cross-inference on the test set with random and unique sampling (Table.~\ref{tab:crossinference}). Random sampling refers to the approach of randomly selecting three \textit{NRD} segments for a given instance, which may result in duplicate samples. On the contrary, unique sampling ensures that no \textit{NRD} segment across the whole image is sampled more than once, thereby guaranteeing uniqueness. Cross-inference outperformed regular inference in terms of both AUC and accuracy.

\begin{table}
	\centering
	\caption{Performance results for different inference methods. CI refers to cross-inference and TTA refers to test time augment.}
	\label{tab:crossinference}
	\begin{tabular}{lccccr} 
		\hline
		Model & CI & TTA & Sampling & Test Accuracy & Test AUC \\
		\hline
		\multirow{3}{*}{TransientViT} & $\times$ & 1x & $\times$ & 98.66 & 97.88\\
		& \checkmark & 3x & unique & \textbf{98.97} & 98.12\\
		& \checkmark & 6x & random & 98.86 & \textbf{98.27}\\
		
		\hline
	\end{tabular}
	
\end{table}

\section{Conclusion}
In this study, we introduced TransientViT, a novel CNN-ViT hybrid real/bogus transient classification algorithm for KATS. We highlighted the limitations of existing CNN-based methods in describing low-level features and the need to improve the efficiency in capturing both local and global features. TransientViT employs convolutional layers to decrease image resolution and utilizes a hierarchical attention mechanism to model the global feature map. It overcomes the aforementioned limitations by combining the locality of CNNs and the global connectivity of ViTs.

Extensive experiments showcased the superiority of TransientViT over various ViT- and CNN-based backbone architectures. TransientViT achieves an AUC of 0.97 and an accuracy of 99.44\% for the KATS-T 200K test dataset. The ablation studies provided insights into the impact of different input channels, multi-input fusion methods, and cross-inference strategies on model performance. Our proposed adaptive cross-attention mechanism played a crucial role in achieving this impressive performance, effectively fusing spatial and temporal features. The utilization of a voting-based ensemble to combine three inference results further contributed to the reliability and robustness of the TransientViT model predictions. 

Upon the incorporation of TransientViT into the KATS pipeline, the number of transient candidates generated each night was reduced to approximately 1/10, thereby dramatically reducing the requirement for manual inspection in the real/bogus transient classification process. Moreover, the reduction in the number of transient candidates did not compromise the system's ability to accurately detect real transient events. We will continue our work to further reduce FPR in the future. There are defects in most of the images in the dataset (e.g., streaks caused by equatorial mount failure). To some extent, image quality limits the performance of TransientViT. We also plan to further reduce the number of parameters of TransientViT in future studies so as to boost its classification efficiency. TransientViT codes and pre-trained model is open source and available at \href{https://github.com/TimeDevBlocker/TransientViT}{https://github.com/TimeDevBlocker/TransientViT}.

\section{Acknowledgements}

We wish to thank the Xinjiang Astronomical Observatory for providing data storage and hardware support for the Kilodegree Automatic Transient Survey. We also wish to acknowledge Quanzhi Ye for his advice on the development of the classifier and Xing Gao for his contributions to the Kilodegree Automatic Transient Survey.

This study employed Astropy:\footnote{http://www.astropy.org} a community-developed core Python package and an ecosystem of tools and resources for astronomy \citep{astropy:2013, astropy:2018, astropy:2022}.


\bibliographystyle{mnras}
\bibliography{transientvit.bib} 



\bsp	
\label{lastpage}
\end{document}